\documentclass[final]{elsarticle_final}

\usepackage{lineno,hyperref,tabularx}
\modulolinenumbers[5]

\journal{Icarus}

\biboptions{authoryear}

\usepackage{amsmath}
\usepackage{caption}

\makeatletter
\def\@author#1{\g@addto@macro\elsauthors{\normalsize%
    \def\baselinestretch{1}%
    \upshape\authorsep#1\unskip\textsuperscript{%
      \ifx\@fnmark\@empty\else\unskip\sep\@fnmark\let\sep=,\fi
      \ifx\@corref\@empty\else\unskip\sep\@corref\let\sep=,\fi
      }%
    \def\authorsep{\unskip,\space}%
    \global\let\@fnmark\@empty
    \global\let\@corref\@empty  
    \global\let\sep\@empty}%
    \@eadauthor={#1}
}
\makeatother

\begin{document}

\begin{frontmatter}

\title{Eddy Evolution during Large Dust Storms}

\author{Michael Battalio\corref{cor1}}
\ead{michael@battalio.com}
\author{Huiqun Wang}
\address{Smithsonian Astrophysical Observatory, Harvard-Smithsonian Center for Astrophysics\\ 60 Garden Street, MS 50, Cambridge, MA 02138}

\begin{abstract}

The evolution of eddy kinetic energy during the development of large regional dust storms on Mars is investigated using the Mars Analysis Correction Data Assimilation (MACDA) reanalysis product and the dust storm data derived from Mars Global Surveyor Mars Daily Global Maps. Transient eddies in MACDA are decomposed into different components according to their eddy periods: $P\leq1$ sol, $1<P\leq8$ sols, $8<P\leq60$ sols. This paper primarily focuses on the Mars year 24 pre-solstice ``A'' storm that starts with many episodes of frontal/flushing dust storms from the northern hemisphere and attains its maximum global mean opacity after dust expansion in the southern hemisphere. During the development of this storm, the dominant eddies in terms of eddy kinetic energy progress from the $1<P\leq8$ sol eddies in the northern mid/high latitudes to the $P\leq1$ sol eddies (dominated by thermal tides) in the southern mid latitudes, and the $8<P\leq60$ sol eddies show a prominent peak with the increased  global-mean dust opacity. The peaks of the $1<P\leq8$ sol eddies are found to best correlate with the average area of textured frontal/flushing dust storms within 40$^\circ$N -- 60$^\circ$N. The region where the $1<P\leq8$ sol eddies increase the most corresponds to the main flushing channel. The eddy kinetic energy of the $P\leq1$ eddies, dominated by $P$ = 1 and its harmonics, increases with the global mean dust opacity both before and after the winter solstice in Mars year 24. 
The $8<P\leq60$ sol eddies briefly spike during large, regional dust storms but remain weak if dust storm sequences do not lead to a major dust storm.  Zonal wavenumber analysis of eddy kinetic energy shows that the peaks of the $1<P\leq8$ eddies often result from combinations of zonal wavenumbers 1 to 3, while the $P\leq1$ eddies and $8<P\leq60$ sol eddies are each dominated by zonal wavenumber 1.

\end{abstract}


\end{frontmatter}


 \section{Introduction}\label{intro}
 	
	Dust storms are among the defining features of the Martian atmosphere. Large regional and global dust storms (GDS) are observed in the dust storm season ($L_s$ = 150$^{\circ}$ -- 330$^{\circ}$) \citep{Martin1993,Wang2015}. These visually impressive storms are associated with dramatic changes in the thermal structure and circulation of the atmosphere \citep[e.g.,][]{Zurek1992,Barnes2017}. Numerous studies have examined the effects of large dust storms and found strengthened zonal-mean Hadley circulation \citep[e.g.,][]{Haberle1993}, increased midlevel temperature in both hemispheres \citep[e.g.,][]{Kass2016}, and complex changes in thermal tides and stationary waves \citep[e.g.,][]{Wilson1996,Banfield2000,Banfield2003,Guzewich2014}. The systematic daily global mapping of Mars Global Surveyor (MGS) and Mars Reconnaissance Orbiter (MRO) shows that major dust storms in non-GDS years are often initiated as frontal/flushing dust storm sequences that are in turn linked to strong transient eddies near the surface \citep{Wang2003,Hinson2012,Wang2015}.

Atmospheric circulation simulated by Global Circulation Models (GCMs) shows complex changes with time during the evolution of modeled dust storms \citep{Murphy1995,Montabone2005}. It  seems reasonable to hypothesize that different development stages of a large dust storm are characterized by different changes in atmospheric circulation.  In this paper, taking advantage of a Mars atmospheric reanalysis product and a Mars dust storm database, we investigate the temporal and spatial changes in various atmospheric eddies by studying their eddy kinetic energy during the evolution of large dust storms in non-GDS years.   We primarily focus on the pre-solstice dust storm of Mars Year (MY) 24, as the quality of the reanalysis is better than that during the dust storms in subsequent years \citep{Montabone2014, Pankine2015,Pankine2016}, but results for MY 26 are  included when appropriate.  
	
 \section{Methods}\label{methods}

The relationship between dust storms and atmospheric circulation is evaluated by analyzing the kinetic energy of various eddies derived from a reanalysis product along with a dust storm database.  In this section, we first present the compilation of the dust storm database, then explain the analysis of the reanalysis product.  

\subsection{Dust storm identification}

Dust storms are identified in Mars Daily Global Maps (MDGMs) that are composed from the wide-angle global map swaths taken by the MGS Mars Orbiter Camera (MOC) at about 2 PM local time \citep{Wang2002}.  Each MDGM (0.1$^\circ$ longitude $\times$ 0.1$^\circ$ latitude) includes a north polar map (45$^\circ$N -- 90$^\circ$N), a south polar map (45$^\circ$S -- 90$^\circ$S), and a non-polar map (60$^\circ$S -- 60$^\circ$N). The MOC MDGM dataset spans from MY 24 $L_s$ = 150$^{\circ}$ to MY 28 $L_s$ = 121$^{\circ}$.  A GDS occurred in MY 25. To focus on non-GDS conditions, the non-polar maps for MY 24 and MY 26 are used to identify dust storms in this paper.

The dust storm identification procedure follows the method used in \cite{Battalio2019}. Briefly, dust storm boundaries are outlined manually in each MDGM and given a confidence level to indicate the accuracy of the storm edge relative to the total storm area. The confidence level varies from 1 to 4 (25$\%$ to 100$\%$), with 4 being the most confident. In this paper, the storms with a confidence level of $\geq$3 are analyzed, and they account for 77\% of all the identified dust storms. Repeating the analysis with less stringent confidence intervals does not substantively alter the results.  Our independently identified dust storms conform to those of \cite{Guzewich2015} and \cite{Kulowski2017}. When dust storms show organized activity, they are grouped into dust storm sequences.  A sequence is a collection of one or more dust storms that are distinct from the background dust opacity and follow a repeatable, coherent general trajectory \citep{Wang2015, Battalio2019}. The lifetime of a sequence is 3 sols or more and is longer than the lifetimes of the dust storm members of the sequence.   A dust storm member can last from $<$1 sols to several sols.  A movie showing the sequences of dust storms contributing to the  MY 24 pre-solstice large, regional dust storm is included in the supplementary material, including 5 sols before and after the event. The dust storm data gathered for this paper can be found in the supplementary material.

\subsection{Kinetic energy analysis}

To examine how the circulation changes with the development of large dust storms, we use the Mars Analysis Correction Data Assimilation (MACDA) v1.0 \citep{Montabone2014} to calculate the kinetic energy ($KE$) of eddies with different oscillation periods ($P$).  The MACDA (5$^\circ$ longitude $\times$ 5$^\circ$ latitude $\times$ 25 level $\times$ 2 hour) is a reanalysis of the MGS Thermal Emission Spectrometer (TES) temperature and dust opacity data using the LMD Mars GCM and has been used to investigate the climatology and structure of transient eddies and thermal tides in the Martian atmosphere \citep[e.g.,][]{Lewis2005,Lewis2016,Battalio2016,Battalio2018}.
The $u$ and $v$ wind from MACDA are each decomposed into four eddy components ($u'$ and $v'$) based on their oscillation periods:  $P\leq1$ sol, $1<P\leq8$ sols, $8<P\leq60$ sols, and $P>60$ sols (quasi-stationary). 
The $P\leq1$ sol eddies include not only contributions from forced thermal tides of $P=1$ sol and its harmonics but also contributions from transient eddies whose periods are less than a sol, but  as will be shown later, the $P\leq1$ sol eddies are dominated by tidal modes.  The KE of each eddy component is calculated at each grid point using the corresponding wind perturbations ($KE = (u'^2+v'^2)/2$). When planetary waves are investigated, we apply a zonal Fourier filter on the corresponding $u'$ and $v'$ to isolate the contributions from each zonal wavenumber and use the resulting perturbations in the calculation. 

The eddy winds are obtained from the following procedure. First, a 60-sol running mean is subtracted from the MACDA time series to generate time perturbations at each grid point. Quasi-stationary ($P>60$ sols) eddies are found by subtracting the zonal mean from the 60-sol running mean.  The other three eddy components ($P\leq1$ sol, $1<P\leq8$ sols, and $8<P\leq60$ sols) are found by filtering the time perturbations using three separate 100th order Hamming windows--one for each component.  Varying the period of the 8-sol cutoff from 5 to 12 sols does not substantively change the results presented in this paper. The Hamming-window band-pass filter (designed using the Matlab function \texttt{fir1}) provides at least 50 dB attenuation in the stopbands and at most a 0.01 dB ripple in the passband.  An example of the eddy time series from the procedure (for 62.5$^\circ$N, 30$^\circ$W, and the $\sigma$=0.9426 vertical level) can be found in Supplementary Fig. 1. 

The total $KE$ above a surface grid point (J/m$^2$) at a given time step is calculated as
\begin{equation}\label{eq:KE}
\overline{KE_{t,\lambda,\theta}} = \frac{1}{2g} \int_{p_t}^{p_s} (u'^2+v'^2) dp,
\end{equation}		
 where $p_s$ is the pressure at the surface, $p_t = p_s * 1.5\times10^{-3}$ is the pressure at the top of the reanalysis, $g=3.71$ m/s$^{2}$, t is time, $\lambda$ is longitude, and $\theta$ is latitude. 
 The global-mean eddy kinetic energy, $\langle KE \rangle$, (J/m$^2$) at each time step is calculated by averaging $\overline{KE_{t,\lambda,\theta}}$ over the globe 
\begin{equation}\label{eq:KEt}
\langle KE \rangle= \frac{1}{4\pi} \int_{\theta=-\pi/2}^{\pi/2} {\cos{\theta}}\int_{\lambda=0}^{2\pi} \overline{KE_{t,\lambda,\theta}}d\lambda d\theta.
\end{equation}		
The compact form of $\langle KE \rangle$ facilitates our discussion of eddy evolution during large dust storms.  

While most results in this paper are based on the definitions above, occasionally, we comment on the results obtained using Eq. \ref{eq:KEnor}--a normalized version of Eq. \ref{eq:KE} that decouples the influence of surface pressure, 
\begin{equation}\label{eq:KEnor}
\overline{KE_{t,\lambda,\theta}}_{norm} = \frac{1}{2(p_s - p_t)} \int_{p_t}^{p_s} (u'^2+v'^2) dp.
\end{equation}		

 \section{Results}\label{results}
 
 We describe the latitudinal and temporal distribution of dust storms in Section \ref{distribution_sec}, evaluate the evolution of global averages of $\langle KE \rangle$ in the context of large dust storms  in Section \ref{evolution_sec}, separate the responses of different zonal wavenumbers for $\langle KE \rangle$ in Section \ref{wavenumber_sec}, and  compare the spatial distribution of vertically integrated $\overline{KE_{t,\lambda,\theta}}$ with that of dust storm activity in Section \ref{spatial_sec}.
 
 \subsection{Dust storm distribution}\label{distribution_sec}
 
\begin{figure}[tph]
  \noindent\includegraphics[width=29pc,angle=0]{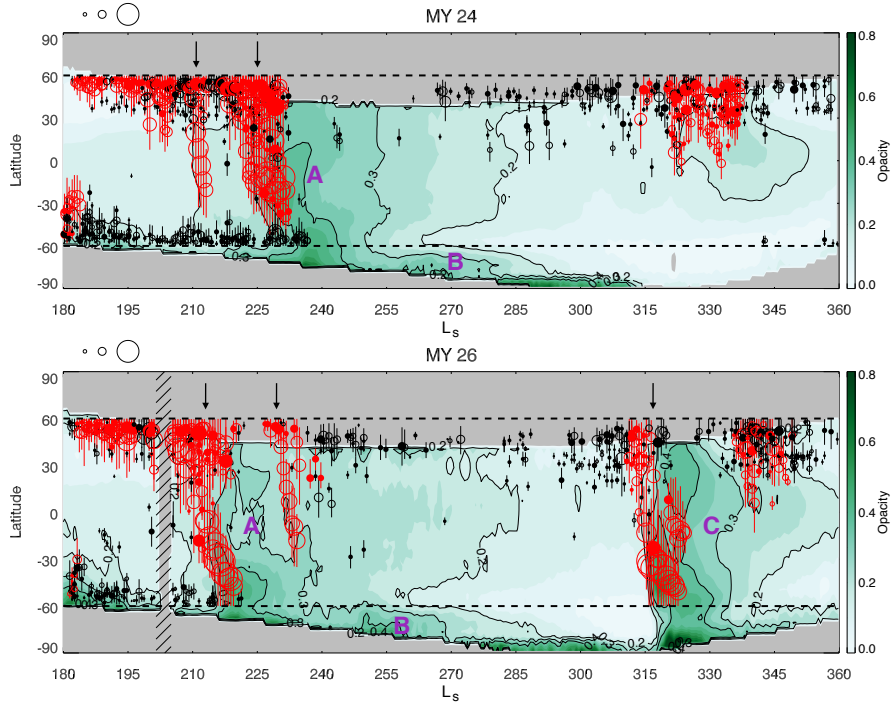}
  \caption{Dust storm distribution for MY 24 (top) and 26 (bottom) derived from the MGS MOC non-polar MDGMs (60$^\circ$S -- 60$^\circ$N, indicated by dashed lines).  Circles correspond to centroid latitudes and $L_s$ values of dust storms and are scaled by dust storm sizes. Filled circles indicate new dust storms.  Legends at the top left of each panel indicate the storm sizes of  $10^6$, $5\times10^6$, and $10^7$ km$^2.$ Vertical lines over the circles indicate the latitudinal extents of dust storms. Dust storms associated with dust storm sequences are highlighted in red; other dust storms are in black. The 9.3 $\mu$m zonal-mean dust opacity (scaled to 610 Pa) \citep{Montabone2015} is color contoured in green. ``A,'' ``B,'' and ``C'' storms as classified by \cite{Kass2016} are indicated. Arrows point to the events discussed in the text.}\label{distribution}
\end{figure}

Figure \ref{distribution} shows the $L_s$ versus latitude distribution of the dust opacity (filled contours) and dust storms (symbols) during $L_s$ = 180$^{\circ}$ -- 360$^{\circ}$ for MY 24 (top) and MY 26 (bottom), respectively.  The 9.3 $\mu$m  dust opacity is the zonal average of the \cite{Montabone2015} gridded product (scaled to 610 Pa). The dust storms identified in MDGMs are indicated with circles scaled by dust storm area. Filled circles indicate new activity, while open circles indicate storms continuing from the previous sol.  The dust storms that are involved in dust storm sequences are highlighted in red, and non-sequence storms are in black.  Organized dust storm sequences are particularly important in the Martian dust cycle, as their members are generally larger than isolated dust storms, and they can lead to major dust storms of global impact \citep{Wang2015,Battalio2019}. In both Mars years, dust storm sequences flush from the northern to the southern hemisphere through longitudinally confined channels \citep{Wang2015}, leading to apparent increases in dust opacity on planetary scale (Fig. \ref{distribution}) \citep{Montabone2015}. Following each burst of activity from a dust storm sequence in the northern high latitudes (e.g., $L_s$ = 210$^{\circ}$ and 225$^{\circ}$  in MY 24 and $L_s$ = 215$^{\circ}$, 230$^{\circ}$, and 315$^{\circ}$ in MY 26, indicated by arrows in Fig. \ref{distribution}), there is a period of reduced dust activity for at least 20 sols.  The quiet period following the MY 24 large dust storm is long, as the pre-solstice storm directly precedes the solstitial pause described in \cite{Lewis2016}. 

\cite{Kass2016} classified dust storms into pre-solstice ``A,'' near-solstice ``B,'' and post-solstice ``C'' storms according to dust opacity and atmospheric thermal signatures observed in MCS retrievals. The ``A'' and ``C'' storms have global impact on atmospheric opacity and circulation, while the ``B'' storms only affect the southern high latitudes.  Figure \ref{distribution} shows that the peak of the MY 24 ``A'' storm is immediately preceded by large flushing dust storm sequences.  These sequences travel through Acidalia, Arcadia, and Utopia, with Acidalia and Arcadia being the main tracks (Fig. \ref{evolution}b) \citep{Wang2015}.  In MY 26, a flushing dust storm sequence from Utopia leads to the MY 26 ``A'' storm around $L_s$ $\sim$ 215$^{\circ}$, and a sequence from Acidalia leads to the MY 26 ``C'' storm around $L_s$ $\sim$ 320$^{\circ}$ (Fig. \ref{evolution}d) \citep{Wang2007}.  MY 24 has a resumption of enhanced dust storm activity after the winter solstice, but the corresponding enhancement of dust opacity and mid-level (50 Pa) atmospheric temperature are not large enough to qualify it as a ``C'' storm according to \cite{Kass2016}. 

The dust storm area identified in MDGMs and the globally averaged dust opacity for MY 24 and 26 are plotted as a function of $L_s$ in Fig. \ref{evolution}b\&d. The duration and size of the dust storm sequences contributing to the dust storm area are indicated using horizontal bars, with different colors representing different origination regions and bar thickness indicating maximum area. Note that dust haze without a clear boundary is not identified as a dust storm in this study, regardless of its optical depth. Thus, there is a sharp drop in dust storm area when active dust lifting centers cease, even though the background dust continues to increase as a result of the previously lifted dust being dispersed by the circulation over planetary scale.  Figure \ref{evolution}b\&d shows that the peak global dust opacity follows the large dust storm sequences, again suggesting that the dust storms within the sequences result in the wide-spread dust haze that shrouds the planet during major dust events.

 \subsection{Evolution of globally averaged eddy kinetic energy}\label{evolution_sec}
	
\begin{figure}[tph]
  \noindent\includegraphics[width=26pc,angle=0]{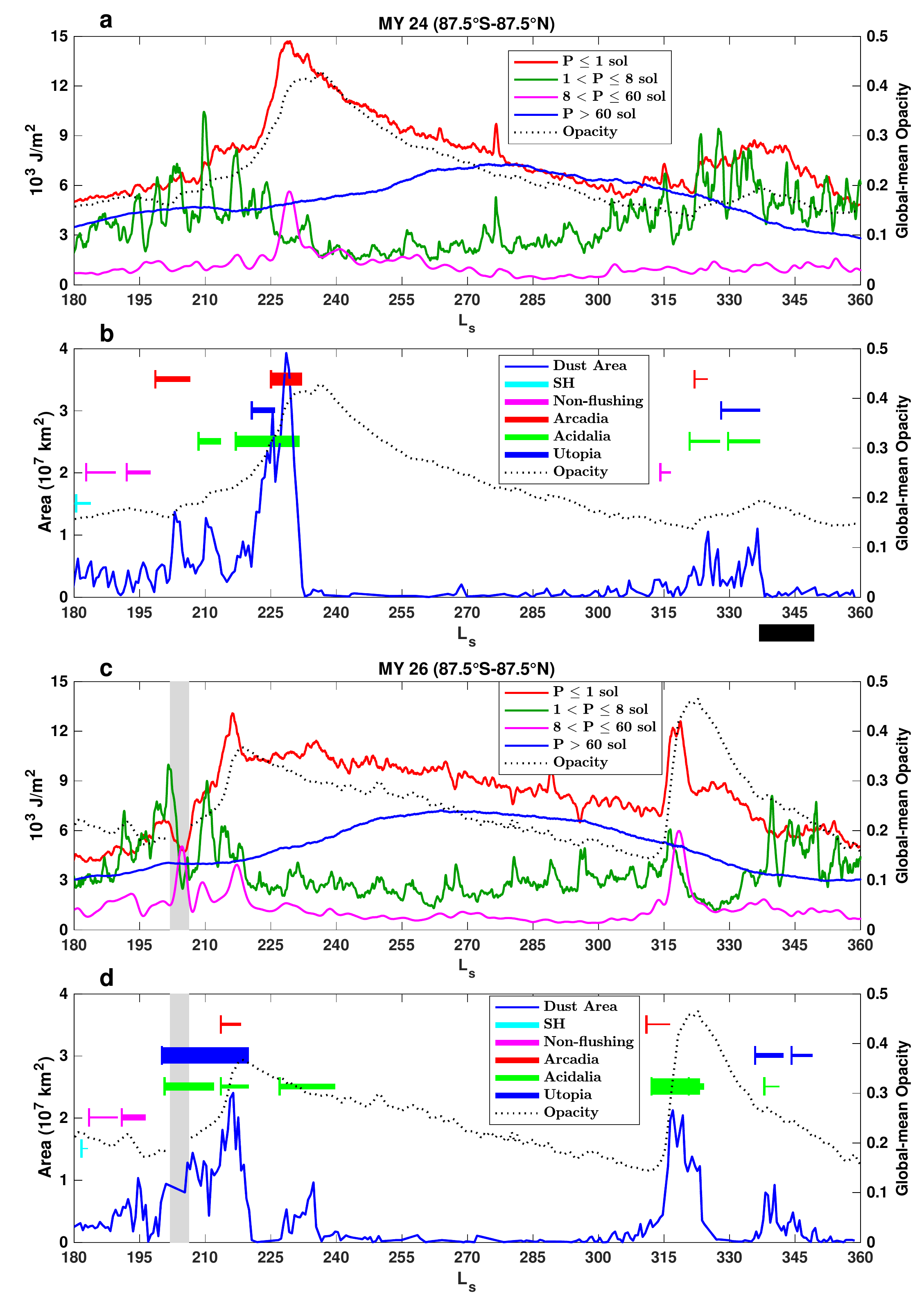}
  \caption{Globally averaged dust opacity (scaled to 610 Pa) (right axis, black) and globally averaged eddy kinetic energy, $\langle KE \rangle$, (left axis) for MY 24 (a) and MY 26 (c).  The curve for the $P\leq1$ sol eddies is smoothed with a 1-sol running mean.   b) and d) Total area of dust storms identified in non-polar MDGMs (left axis, blue solid line) and globally averaged dust opacity (scaled to 610 Pa, right axis, black dotted line).  Dust storm sequences are color coded according to their origination areas, with the bar thickness scaled to the maximum area of the sequence.  ``SH'' refers to southern hemisphere sequences.  The gray vertical bar denotes the time period when MACDA is unconstrained due to lack of MGS TES data.  The black bar below panel (b) indicates a sequence area of 2 $\times$ 10$^7$ km$^2$.}\label{evolution}
\end{figure}

	Figure \ref{evolution}a\&c shows the time-series of globally averaged eddy kinetic energy, $\langle KE \rangle$, for each eddy component during $L_s$ = 180$^{\circ}$ -- 360$^{\circ}$ for MY 24 and 26.   The curve for the $P\leq1$ sol eddies is smoothed using a 1-sol running window for ease of comparison with the results for other eddies. As a sanity check, the multi-year (MY 24 -- MY 26)  average of the sum of the $P\leq1$ sol, $1<P\leq8$ sol, and $8<P\leq60$  sol eddy $\langle KE \rangle$ is identical to that reported by \cite{Tabataba-Vakili2015}:  13 $\times$ 10$^3$ J/m$^{2}$. The global-mean dust opacity derived from the \cite{Montabone2015} gridded product (black dashed line, right axis) is over-plotted for reference, and the largest three peaks correspond to the ``A'' and ``C'' storms in MY 24 and 26 (Fig. \ref{distribution}). After the pre-solstice ``A'' storms, the global dust opacity declines gradually through the solstice period until it is briefly elevated by the post-solstice dust storm sequences.

The $P\leq1$ sol eddies (red lines in Fig. \ref{evolution}a\&c) usually have the largest eddy $\langle KE \rangle$. 
The seasonal evolution of $P\leq1$ sol eddies follows that of the globally averaged dust opacity. Thus, the $P\leq1$ sol eddies are good indicators of the dust loading over the planet, as also found by previous studies using other meteorological fields \citep{Banfield2003,Lewis2005}.  Based on MACDA, the $P=1$ sol eddies account for 83$\%$ of the power of the $P\leq1$ sol eddies in terms of $\langle KE \rangle$, and $P=0.5$ sol eddies account for 9$\%$.    
Our results conform to previous findings that $P\leq1$ sol eddies, particularly the diurnal and semi-diurnal thermal tides, are greatly amplified during large dust storms \citep{Leovy1979,Wilson1996,Banfield2000,Banfield2003}.  However, a caveat should be noted that the analysis of  $P\leq1$ sol eddies in MACDA is hindered by the lack of full local time coverage of TES observations (at $\sim$0300 and $\sim$1500)  and the simple treatment of the vertical dust distribution \citep{Navarro2017}.  Nevertheless, previous work has shown that temperature assimilation in MACDA provides some constraint on the thermally forced tide portion of the  $P\leq1$  sol eddies  \citep{Lewis2005,Ruan2019}, and the results still provide useful information.  

On a closer examination for the ``A'' and ``C'' dust storms in MY 24 and 26, we find that the $P\leq1$ eddies and dust opacity do not follow each other exactly. Specifically, the peak of the $P\leq1$ sol eddies generally lags behind the series of dust storm sequences that contributes to the peak of the dust storm area, but the peak of the $P\leq1$ sol eddies leads slightly ahead of the peak of the globally averaged dust opacity. This suggests that both the excitation by dust storms during the transition from a relatively clear to a dusty atmosphere and the maintenance in a dusty atmosphere are important for the evolution of the $P\leq1$ eddies.  

The $\langle KE \rangle$ of the quasi-stationary eddies (blue lines in Fig. \ref{evolution}a\&c) exhibits a slowly varying seasonal cycle during northern fall and winter. Larger values are found during the winter solstice period when they are sometimes of comparable magnitude to the $P\leq1$ eddies. Such a seasonal cycle is consistent with that of stationary waves derived from MGS TES data \citep{Banfield2003}. It is in contrast to the pattern exhibited by the $P\leq1$ sol eddies. While the $P\leq1$ sol eddies are strongly controlled by dust activity, the quasi-stationary eddies vary gradually and appear largely insensitive to dust activity for MY 24 and 26. 
The slope of the curves might suggest that quasi-stationary eddies are slightly curtailed after the MY 24 an MY 26 ``A'' storms, but it is indecisive.  It is also not clear if the MY 26 ``C'' storm has any effect on these eddies.  Using TES data during the Aerobreaking and Science Phasing period of MGS (MY 23), \cite{Banfield2003} reported that the southern hemisphere stationary waves show little change during the pre-solstice ($L_s$ = 225$^{\circ}$ -- 240$^{\circ}$) ``A'' storm but substantial enhancement during the post-solstice ($L_s$ = 310$^{\circ}$ -- 320$^{\circ}$) ``C'' storm that year. They suggested that the difference in stationary waves was due to the difference in the zonal distribution of dust between the two dust storms.  As stationary eddies do not show consistent correlation with large dust storms, we do not examine them further in this paper. 

The $\langle KE \rangle$ for the $1<P\leq8$ sol eddies (green lines in Fig. \ref{evolution}a\&c) has multiple peaks approaching or exceeding 6 $\times$ 10$^3$ J/m$^{2}$ during $L_s$ = 195$^{\circ}$ -- 225$^{\circ}$, before the MY 24 ``A'' storm. Each peak corresponds to a burst of dust activity associated with one or more dust storm sequences. After $L_s$ = 225$^{\circ}$, the $\langle KE \rangle$ drops down to a lower level of around 3 $\times$ 10$^3$ J/m$^{2}$. During $L_s$ = 195$^{\circ}$ -- 225$^{\circ}$, the strength of the $1<P\leq8$ sol eddies rivals that of the $P\leq1$ sol eddies and even briefly surpasses all the other eddies. The largest peak of the $1<P\leq8$ sol eddies is at $L_s$ $\sim$ 210$^{\circ}$ and coincides with an Acidalia dust storm sequence that involves a series of flushing dust storms occurring on a daily basis (Fig. \ref{evolution}b). The flushing storms crossed the equator and reached the area between Valles Marineris and Argyre but showed little zonal expansion. This event is accompanied by a noticeable (albeit small) increase in global-mean dust opacity and a corresponding increase in the $\langle KE \rangle$ of $P\leq1$ eddies. Following a 9-sol hiatus, the $1<P\leq8$ sol eddies return.   Dust storm sequences also resume in Acidalia, Utopia, and Arcadia (Fig. \ref{evolution}b), eventually leading to the MY 24 ``A'' storm.  This conforms to the idea that $1<P\leq8$ sol eddies play an important role in initiating/supporting frontal/flushing dust storms \citep{Wang2003}. With the increase of the background dust opacity and the $P\leq1$ sol eddies  toward their peaks for the MY 24 ``A'' storm, the $1<P\leq8$ sol eddies decrease. This suggests a negative feedback, whereby stabilization of the synoptic weather disturbances occurs under increased background dust \citep{Murphy1995,Battalio2016,Lee2018}.  After the winter solstitial pause \citep{Lewis2016}, dust storm sequences return along with strong $1<P\leq8$ sol eddies, though the post-solstice dust storm sequences do not lead to a ``C'' storm in MY 24. Thus, it seems that strong synoptic eddies facilitate frontal/flushing dust storms but do not guarantee a major dust storm.        

The general patterns exhibited by the MY 24 ``A'' storm are echoed by the MY 26 ``A'' storm, despite a period of missing data during $L_s$ = 202$^{\circ}$ -- 206$^{\circ}$ in MY 26.  The $1<P\leq8$ sol synoptic period eddies are strong when dust storm sequences are active during $L_s$ = 200$^{\circ}$ -- 220$^{\circ}$.  (The MACDA results within the missing data period, indicated by gray vertical bars in Fig. \ref{evolution}c\&d, should be discounted as they are unconstrained by observations.) These dust storm sequences lead to the MY 26 ``A'' storm primarily via the Utopia flushing channel (Fig. \ref{evolution}d). The $1<P\leq8$ sol eddies are suppressed when the global dust opacities increase significantly.  The strength of the synoptic eddies in MACDA is likely affected by the polar temperature error during the fall and winter of MY 26 \citep{Pankine2015,Pankine2016}; nevertheless, the return of substantial $1<P\leq8$ sol eddies after the solstitial pause is accompanied by dust storm sequences.  The $L_s$ = 310$^{\circ}$ -- 320$^{\circ}$ dust storm sequence through Acidalia leads to the MY 26 ``C'' storm, whose peak opacity of 0.47 occurs around $L_s$ = 320$^{\circ}$ (Fig. \ref{evolution}d). The $1<P\leq8$ sol synoptic eddies are suppressed after the peak of the MY 26 ``C'' storm (Fig. \ref{evolution}c). Unlike the MY 24 ``A'' storm which decays slowly, the MY 26 ``C'' storm quickly subsides. By $L_s$ = 340$^{\circ}$, the global-mean dust opacity decreases to 0.3, at which point, the $1<P\leq8$ sol synoptic eddies and dust storm sequences return once more (Fig. \ref{evolution}c\&d).  

The $8<P\leq60$ sol eddies (magenta lines in Fig. \ref{evolution}a\&c) are typically the weakest. However, they show prominent peaks during the ``A'' and ``C'' storms.  When the peaks of the $8<P\leq60$ sol eddies occur, they consistently lag behind those of the $1<P\leq8$ sol eddies and appear close in time to the peaks of the $P\leq1$ eddies. Notice that the $P\leq1$ sol eddies show a weak, broad peak during $L_s$ = 320$^{\circ}$ -- 345$^{\circ}$ in MY 24, but the $8<P\leq60$ sol eddies stay weak throughout that winter. Intriguingly, although the post-solstice dust storm sequences in MY 24 lead to a small increase in global dust opacity and some enhancement of the $P\leq1$ sol eddies, no ``C'' storm occurs that year.  
Thus, the $8<P\leq60$  sol eddies are closely related to mature, large dust storms and not all dust storm sequences.
\cite{Wang2017} found that westward traveling waves are among the signatures of major dust storms on Mars. Those traveling waves are characterized by wave periods longer than about 15 sols and are included in the $8<P\leq60$ sol eddies shown in Fig. \ref{evolution}.  It should be noted that despite the  narrowness of the peaks of the $8<P\leq60$  sol  eddies in Fig. \ref{evolution}a\&c, these results are consistent with \cite{Wang2017} because the vertical integration in the $\langle KE \rangle$ calculation and the non-linear relationship between $\langle KE \rangle$ and the eddy winds sharpens the changes in $\langle KE \rangle$ versus the signatures in temperature observations.  \cite{Murphy1995} also found a slowly propagating westward wave in their global dust storm simulation and suggested that the westward wave signals polar warming.

The results above are for the overall changes of the global average of $\overline{KE_{t,\lambda,\theta}}$,  which include the contribution from surface pressure variations. To test the influence of surface pressure, we use Equation \ref{eq:KEnor} and \ref{eq:KEt} to calculate the global average of normalized eddy $\overline{KE_{t,\lambda,\theta}}_{norm}$ (J/kg). Results are shown in Supplementary Fig. 2.  In comparison with the non-normalized results of Fig, \ref{evolution}, the amplitudes of the $P\leq1$ sol eddies appear larger than the amplitudes of the other eddies. However, the main conclusions about the lead/lag relationship and relative amplitudes among the eddies  remain the same.
	
 \subsection{Zonal wavenumber analysis}\label{wavenumber_sec}
 
\begin{figure}[tph]
  \noindent\includegraphics[width=29pc,angle=0]{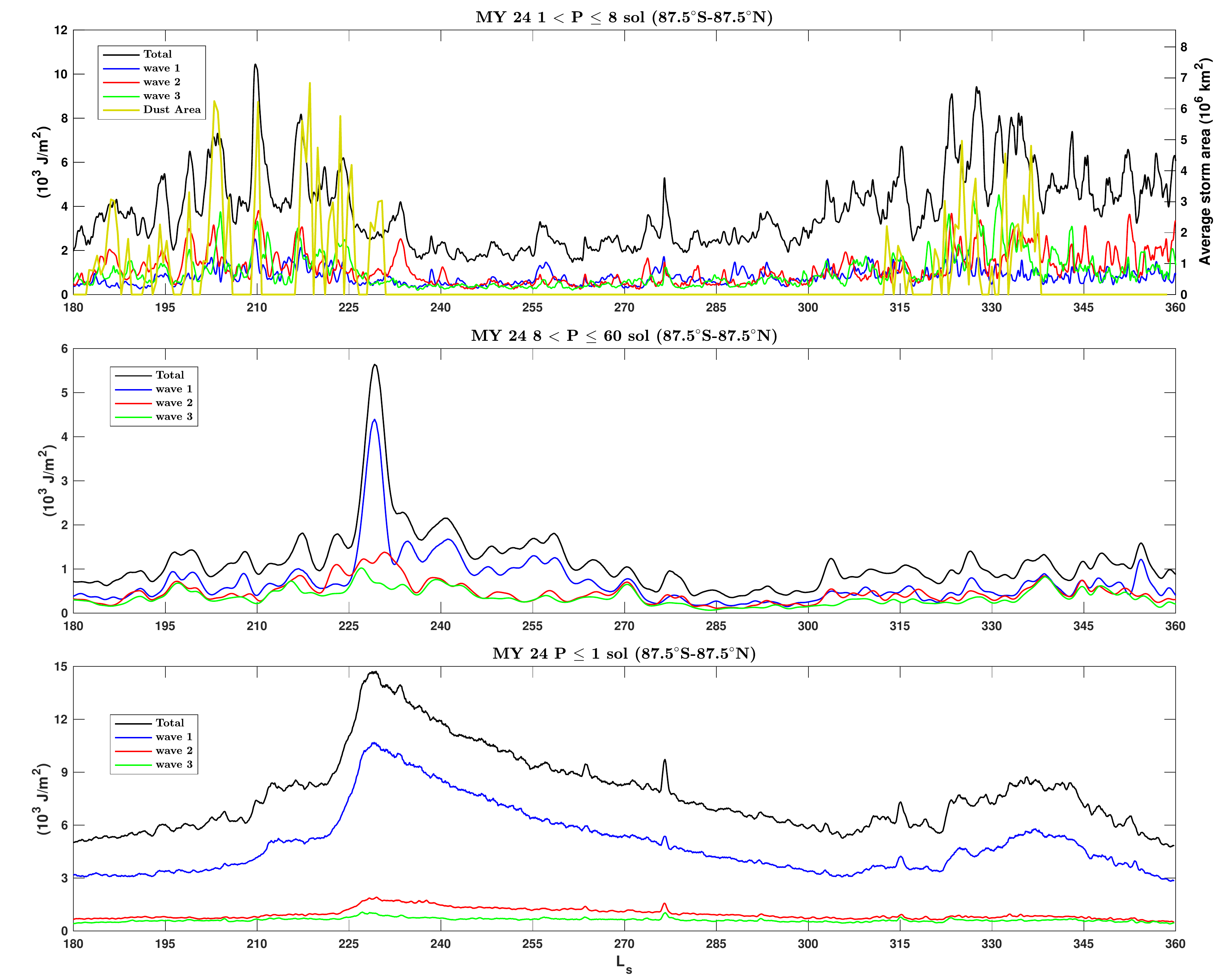}
  \caption{Globally averaged eddy kinetic energy, $\langle KE \rangle$, for $1<P\leq8$ sol (a), $8<P\leq60$ sol (b), and  $P\leq1$ sol eddies (c) for MY 24, filtered by zonal wavenumbers, represented by different colors.  The averaged area of textured, frontal storms within 40$^\circ$N -- 60$^\circ$N is superimposed as a gold line in the top panel (right axis).}\label{wavenumber}
\end{figure}

\begin{figure}[tph]
  \noindent\includegraphics[width=29pc,angle=0]{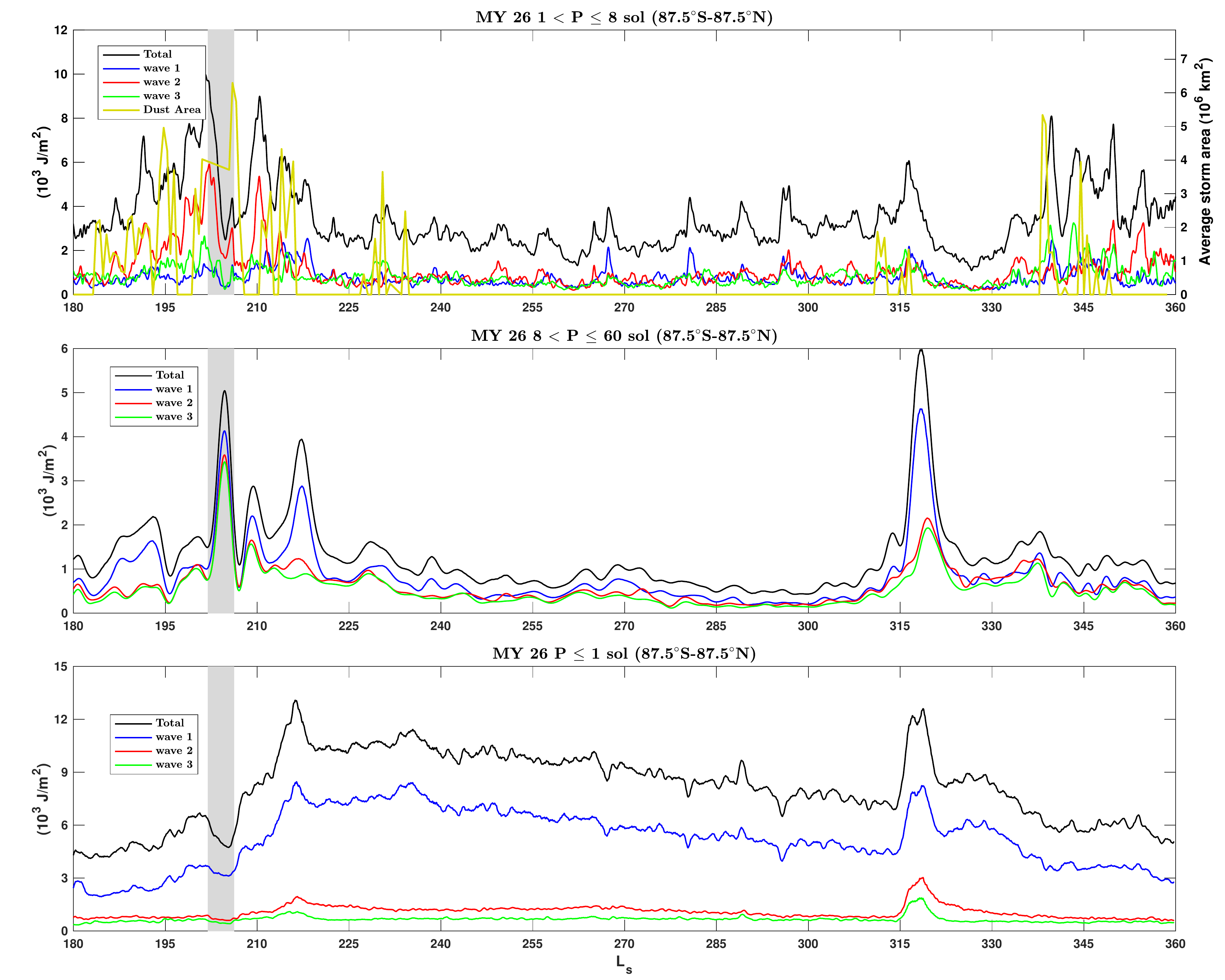}
  \caption{As in Fig. \ref{wavenumber}, but for MY 26. The gray vertical bars denote the time period when MACDA is unconstrained due to lack of MGS TES data.}\label{wavenumber26}
\end{figure}

 The transient eddies in Fig. \ref{evolution} are further examined for different zonal wavenumbers ($k$). 
 Figures \ref{wavenumber} (MY 24) and  \ref{wavenumber26} (MY 26) show the $\langle KE \rangle$ of the $k$ = 1 -- 3 waves  as a function of $L_s$. The $\langle KE \rangle$ of other zonal wavenumbers is far smaller and therefore not shown. The peaks of the $1<P\leq8$ sol eddies typically align with the peaks of $k$ = 1 -- 3 waves (Figs. \ref{wavenumber}a and \ref{wavenumber26}a). Sometimes, a single wavenumber dominates, for example in MY 24 $k$ = 3 at $L_s$ = 225$^{\circ}$ and 330$^{\circ}$ 
 or in MY 26 $k$ = 2 at $L_s$ = 211$^{\circ}$; 
 more frequently, a combination of different zonal wavenumbers contributes (e.g., $k$ = 1 -- 3 at $L_s$ = 195$^{\circ}$,  210$^{\circ}$, 217$^{\circ}$, 323$^{\circ}$, and 328$^{\circ}$ in MY 24 or  $k$ = 1 -- 3 at $L_s$ = 340$^{\circ}$ in MY 26). Interaction of multiple wave modes can lead to sharpened gradients or enhanced variability in certain locations that may be inductive to frontal/flushing dust storms \citep{Banfield2004}.

The dust storms identified in MDGMs have a variety of morphologies and shapes. To illustrate the relationship between dust storms and transient eddies, we tested different filtering criteria to see which subset of dust storms correlated the best with the $1<P\leq8$ sol eddies. The subset is found to be the average area of textured frontal/flushing dust storms within 40$^\circ$N -- 60$^\circ$N (Figs. \ref{wavenumber}a and \ref{wavenumber26}a, gold line right axis). Note, as the MDGMs used in this study cover 60$^\circ$S -- 60$^\circ$N only, high latitude dust storms are not observed, though they should conceivably contribute to dust storm area. Nevertheless, the average area of the selected dust storms correlates reasonably with the $\langle KE \rangle$ of the $1<P\leq8$ sol eddies, with a linear correlation coefficient of $r$ = 0.53 for MY 24.  The pre-solstice period ($L_s$ = 180$^{\circ}$ -- 270$^{\circ}$) has a larger correlation coefficient ($r$ = 0.66) than the post-solstice period ($L_s$ = 270$^{\circ}$ -- 360$^{\circ}$, $r$ = 0.54).   Particularly, the peaks of $\langle KE \rangle$ at $L_s$ = 194$^{\circ}$, 199$^{\circ}$, 203$^{\circ}$, 210$^{\circ}$, 217$^{\circ}$, 327$^{\circ}$, and 331$^{\circ}$ in MY 24 are coincident with the peaks in the areas of those dust storms (Fig \ref{evolution}a). The correspondence can also be seen in Supplementary Fig. 3 where the KE of the $1<P\leq8$ sol eddies averaged from 0 to 20 km and across 40$^\circ$N -- 60$^\circ$N is plotted instead.   A similar pattern of correlation between average dust storm area and $1<P\leq8$ sol eddy $\langle KE \rangle$ holds for MY 26 (Fig. \ref{wavenumber26}a), with correlation coefficients of $r$ = 0.46 for $L_s$ = 180$^{\circ}$ -- 360$^{\circ}$, $r$ = 0.52 for $L_s$ = 180$^{\circ}$ -- 270$^{\circ}$, and $r$ = 0.32 for $L_s$ = 270$^{\circ}$ -- 360$^{\circ}$.  Textures in dust storms were previously used as proxies for active dust lifting \citep{Guzewich2015}, though it takes some time for textures generated by fresh lifting to dissipate.  The correlation between textured dust storms and the $1<P\leq8$  sol eddies lends some support to the assumption that strong winds associated with $1<P\leq8$ sol eddies facilitate dust lifting and/or development of frontal dust storms.

Previous studies found that frontal/flushing dust storms often occur when $k$ = 3 traveling waves are present either by themselves or in combination with other zonal wavenumbers \citep{Wang2007,Hinson2012,Wang2018}. This tendency is reflected in Fig. \ref{wavenumber}a. Note that flushing storms can occur when $k$ = 3 is not dominant and, in rare cases, not even present \citep{Greybush2019}. The strongest peak of the $1<P\leq8$ sol eddies in the fall of MY 24 occurs at $L_s$ = 210$^{\circ}$ when all three zonal wavenumbers are substantial. The $\langle KE \rangle$ of the $k$ = 2 traveling waves is slightly stronger than that of the $k$ = 3 waves in Fig. \ref{wavenumber}a, though it should be noted that the $\langle KE \rangle$ in this plot considers the whole atmospheric column, while the $k$ = 3 waves are vertically confined closer to the surface than smaller zonal wavenumbers \citep{Banfield2004,Lewis2016}. While the role of $k$ = 3 waves should be acknowledged, Figs. \ref{wavenumber}a\&\ref{wavenumber26}a suggest that the increase of multiple zonal wavenumbers is frequently responsible for local maxima of the $\langle KE \rangle$ of $1<P\leq8$ sol eddies. Although a flushing dust storm sequence occurs near $L_s$ = 210$^{\circ}$ in MY 24, the largest flushing dust storm sequence that year is observed during $L_s$ = 220$^{\circ}$ -- 230$^{\circ}$ (Figs. \ref{distribution}\&\ref{evolution}). The $\langle KE \rangle$ of the $1<P\leq8$ sol eddies during $L_s$ = 220$^{\circ}$ -- 230$^{\circ}$ is still high but weaker than that at $L_s$ = 210$^{\circ}$. \cite{Wang2018} argued that the seasonality and latitudinal position of $k$ = 3 traveling waves make it easier for dust in high latitude frontal dust storms to be entrained into the low latitude circulation.
 
Figure \ref{wavenumber}b shows that the $8<P\leq60$ sol eddies are dominated by $k$ = 1 during $L_s$ = 225$^{\circ}$ -- 260$^{\circ}$, a time period that includes the MY 24 ``A'' storm.  The $k$ = 2 waves appear to be slightly enhanced during $L_s$ = 220$^{\circ}$ -- 235$^{\circ}$. Larger wavenumbers remain weak throughout the fall and winter.  A similar pattern  occurs in MY 26 (Fig. \ref{wavenumber26}b) for the ``A'' storm ($L_s$ = 215$^{\circ}$) and the ``C'' storm ($L_s$ = 320$^{\circ}$).  The peak across all wavenumbers in MY 26 near $L_s$ = 202$^{\circ}$ should be discounted due to the model being unconstrained by observations at the time.  Figures \ref{wavenumber}b\&\ref{wavenumber26}b are consistent with TES observations that traveling waves with long wave periods are mostly $k$ = 1 \citep{Banfield2004}.

The $P\leq1$ sol eddies are dominated by $k$ = 1 waves (Figs. \ref{wavenumber}c\&\ref{wavenumber26}c). 
The $k$ = 2, $P\leq1$ sol eddies also show an apparent increase near $L_s$ = 225$^{\circ}$ in association with the MY 24 ``A'' storm, mostly due to the enhancement of the migrating semi-diurnal tide. 
For $k$ = 1,  the migrating diurnal tide makes up $\sim$85$\%$ of the $\langle KE \rangle$, and for $k$ = 2, the semi-diurnal tide makes up $\sim$45$\%$.  Previous studies showed that thermal tides, especially the diurnal and semi-diurnal tides, are strongly coupled to large dust storms, though the details of change in each tidal mode are quite complex \citep{Leovy1979,Banfield2003,Lewis2005,Guzewich2014}. While the amplitudes of thermal tides greatly increase during large dust storms, Figs. \ref{wavenumber}c\&\ref{wavenumber26}c show that the percentages of their $\langle KE \rangle$ with respect to the total $\langle KE \rangle$ of the $P\leq1$ sol eddies are more stable. For example, $k$ = 1 accounts for $\sim$60$\%$ of the total $\langle KE \rangle$ outside the MY 24 ``A'' storm and increases slightly to $\sim$70$\%$ during it; $k$ = 2 contributes $\sim$12$\%$ throughout the fall and winter.  As noted in Sec. \ref{evolution_sec}, the results concerning thermal tides should be viewed with the caveat about MACDA's possible limitations in mind.

\subsection{Spatial distribution}\label{spatial_sec}

\begin{figure}[tph]
  \noindent\includegraphics[width=29pc,angle=0]{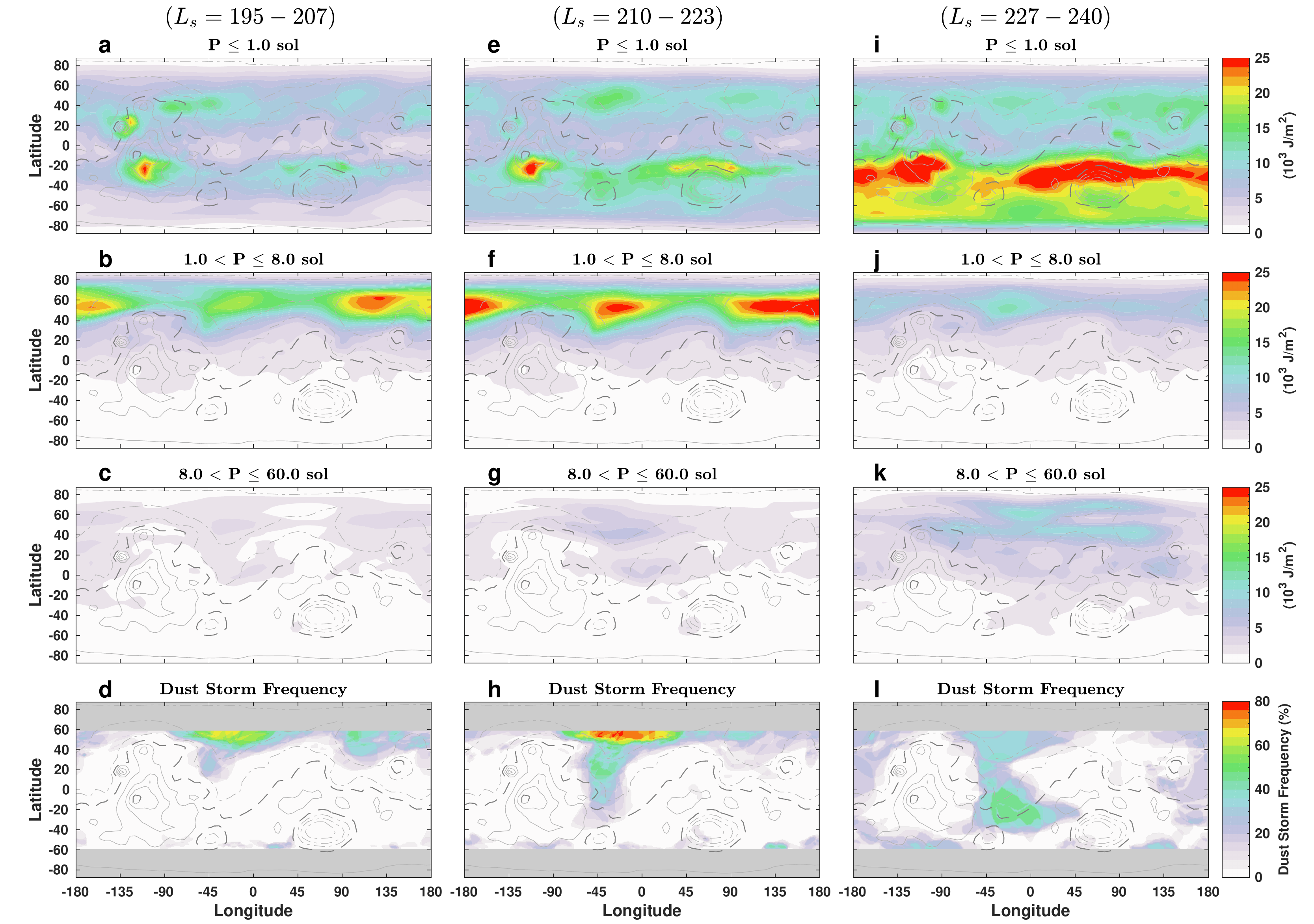}
  \caption{Eddy kinetic energy, $\overline{KE_{t,\lambda,\theta}}$, for $L_s$ = 195$^{\circ}$ -- 207$^{\circ}$ (left column, panels a -- d), $L_s$ = 210$^{\circ}$ -- 223$^{\circ}$ (middle column, panels e -- h), and $L_s$ = 227$^{\circ}$ -- 240$^{\circ}$ (right column, panels i -- l) in MY 24.  The top three rows correspond to $P\leq1$ sol (panels a, e, i), $1<P\leq8$ sol (panels b, f, j), and $8<P\leq60$ sol  (panels c, g, k) eddies, respectively.  The bottom row shows the dust storm occurrence frequency derived from MDGMs for each corresponding time period (panels d, h, l).  Topography is indicated by gray contours, with the thick dashed line at 0 m.}\label{map}
\end{figure}

In Fig. \ref{map}, we investigate the spatial distribution of eddy kinetic energy, $\overline{KE_{t,\lambda,\theta}}$, and dust storm activity for the MY 24 ``A'' storm. We concentrate on $L_s$ = 195$^{\circ}$ -- 207$^{\circ}$, $L_s$ = 210$^{\circ}$ -- 223$^{\circ}$, and $L_s$ = 227$^{\circ}$ -- 240$^{\circ}$ that correspond to the time periods before, immediately preceding, and after the largest combined area of flushing dust storm sequences that lead to the MY 24 ``A'' storm. The choice of the time windows is guided by the timing of changes in eddies.  The dust storm occurrence map for each $L_s$ period is shown in the bottom row of Fig. \ref{map}. The frequency of dust storms at each pixel is calculated as the ratio between the number of times the pixel is identified as a dust storm pixel and the number of times the pixel is observed during the corresponding $L_s$ period. 

During $L_s$ = 195$^{\circ}$ -- 207$^{\circ}$, dust activity is mainly confined north of $\sim$40$^\circ$N, with hotspots in Acidalia and Utopia (Fig. \ref{map}d). This distribution is determined by cap edge dust storms that bend southward in the Acidalia and Utopia storm zones \citep{Hollingsworth1996b}. With the exception of one flushing dust storm from Acidalia to Chryse, all the others remain in the vicinity of the polar cap. During the initiation phase of the MY 24 ``A'' storm ($L_s$ = 210$^{\circ}$ -- 223$^{\circ}$), dust storm occurrence frequency dramatically increases at northern mid/high latitudes, with 6 events developing into flushing dust storms through the Acidalia-Chryse channel (Fig. \ref{map}h). These successive flushing dust storms correspond to the height of the MY 24 flushing activity. They eventually crossed Valles Marineris into the southern hemisphere. The next phase of the MY 24 ``A'' storm ($L_s$ = 227$^{\circ}$ -- 240$^{\circ}$) is characterized by zonal dust expansion from Valles Marineris toward the area northwest of Hellas. The most frequent dust storm occurrence region lies between Valles Marineris and Argyre. The occurrence frequency declines eastward toward the region to the northwest of Hellas (Fig. \ref{map}l).  In addition, this window includes activity from the Arcadia storm track that at its peak area was of similar size to the Acidalia sequence.  However, the Arcadia sequence did not spread as widely in the southern hemisphere as the Acidalia sequence did (Fig. \ref{map}l).

The $\overline{KE_{t,\lambda,\theta}}$ for the $P\leq1$ sol, $1<P\leq8$ sol and $8<P\leq60$ sol eddies is time averaged for each $L_s$ period and plotted in the top three rows of Fig. \ref{map}.  As discussed in Section \ref{evolution_sec}, the $P\leq1$ sol eddies strengthen as the MY 24 ``A'' storm grows. The globally averaged $\overline{KE_{t,\lambda,\theta}}$ (i.e., $\langle KE \rangle$) increases by $\sim$34$\%$ from $L_s$ = 195$^{\circ}$ -- 207$^{\circ}$ (6.09 $\times$ 10$^3$ J/m$^{2}$) to $L_s$ = 210$^{\circ}$ -- 223$^{\circ}$ (8.19 $\times$ 10$^3$ J/m$^{2}$), and increases by another $\sim$63$\%$ from $L_s$ = 210$^{\circ}$ -- 223$^{\circ}$ to $L_s$ = 227$^{\circ}$ -- 240$^{\circ}$ (13.30 $\times$ 10$^3$ J/m$^{2}$).  As the $P\leq1$ sol eddies are dominated by diurnal tides, they straddle the equator with a northern and a southern branch during $L_s$ = 195$^{\circ}$ -- 207$^{\circ}$, as expected from tidal theory and model simulations \citep{Wilson1996, Barnes2017}. Local maxima are seen on the slopes of Hellas and Tharsis, reflecting the effect of slope winds in the regions. The southern branch amplifies dramatically after the flushing sequence progresses to the southern hemisphere, in accordance with an increase in tidal forcing by regional dust loading \citep{Murphy1995,Guzewich2014}. Although the dust occurrence frequency maximizes in the longitudinal sector between Argyre and Hellas, the southern hemisphere $P\leq1$ sol eddies increase at all longitudes with local enhancement near Hellas and Tharsis, indicating the existence of zonally coherent wave modes. In comparison with the southern hemisphere, the average of the northern hemisphere's $\overline{KE_{t,\lambda,\theta}}$ only increases by $\sim$30$\%$ from $L_s$ = 195$^{\circ}$ -- 207$^{\circ}$ to $L_s$ = 227$^{\circ}$ -- 240$^{\circ}$.  

The time evolution of the $P\leq1$ sol eddies can be tied to the development trajectory of the MY 24 ``A'' storm.  Though the primary strengthening of the $P\leq1$ eddies (Fig. \ref{map}e) occurs as dust storms reach the southern hemisphere (Fig. \ref{map}h), this happens after substantial lifting in the northern hemisphere has already occurred before dust storm flushing, hence there is a lag between the peak of dust storm area and the peak of $P\leq1$ sol eddies (Fig. \ref{evolution}a).  Once lifting begins in earnest in the southern hemisphere, the increase in the $\langle KE \rangle$ of the $P\leq1$ sol eddies is so pronounced that its peak occurs before dust is fully distributed around the planet (supplementary material MY24A movie).  

Figure \ref{map}b, f, \& j shows that the $1<P\leq8$ sol eddies first strengthen, then weaken. These eddies are concentrated between 40$^\circ$N and 70$^\circ$N, with local maxima in the Arcadia, Acidalia and Utopia storm zones \citep{Hollingsworth1996b,Banfield2004}. The $\sim$13$\%$ strengthening of the global average of the $1<P\leq8$ sol $\overline{KE_{t,\lambda,\theta}}$ from $L_s$ = 195$^{\circ}$ -- 207$^{\circ}$ (4.72 $\times$ 10$^3$ J/m$^{2}$) to $L_s$ = 210$^{\circ}$ -- 223$^{\circ}$ (5.34 $\times$ 10$^3$ J/m$^{2}$) corresponds to an increase of dust storm occurrence frequency at the polar cap edge and an increase of the number of flushing dust storms. Although the increase in $1<P\leq8$ sol eddies occurs at all longitudes within 40$^\circ$N -- 70$^\circ$N, the most prominent increase occurs in Acidalia, which is the primary flushing channel for the MY 24 ``A'' storm.  The 51$\%$ weakening of the global average of the $1<P\leq8$ sol  $\overline{KE_{t,\lambda,\theta}}$  from $L_s$ = 210$^{\circ}$ -- 223$^{\circ}$ (5.34 $\times$ 10$^3$ J/m$^{2}$) to $L_s$ = 227$^{\circ}$ -- 240$^{\circ}$ (2.61 $\times$ 10$^3$ J/m$^{2}$) in MY 24 corresponds to the cessation of northern hemisphere flushing events and the expansion of southern hemisphere dust that lead to a large increase in the global-mean dust opacity (Fig. \ref{distribution}). The increased dust opacity tends to result in increased atmospheric stability that suppresses baroclinic instability and traveling waves \citep{Battalio2016,Mulholland2016,Lee2018,Greybush2019}.  %
With the development of the MY 24 ``A'' storm, the strongest transient eddies switch from $1<P\leq8$ sol eddies in the northern hemisphere ($L_s$ = 210$^{\circ}$ -- 223$^{\circ}$) to $P\leq1$ sol eddies in the southern hemisphere ($L_s$ = 227$^{\circ}$ -- 240$^{\circ}$).

\begin{figure}[tph]
  \noindent\includegraphics[width=29pc,angle=0]{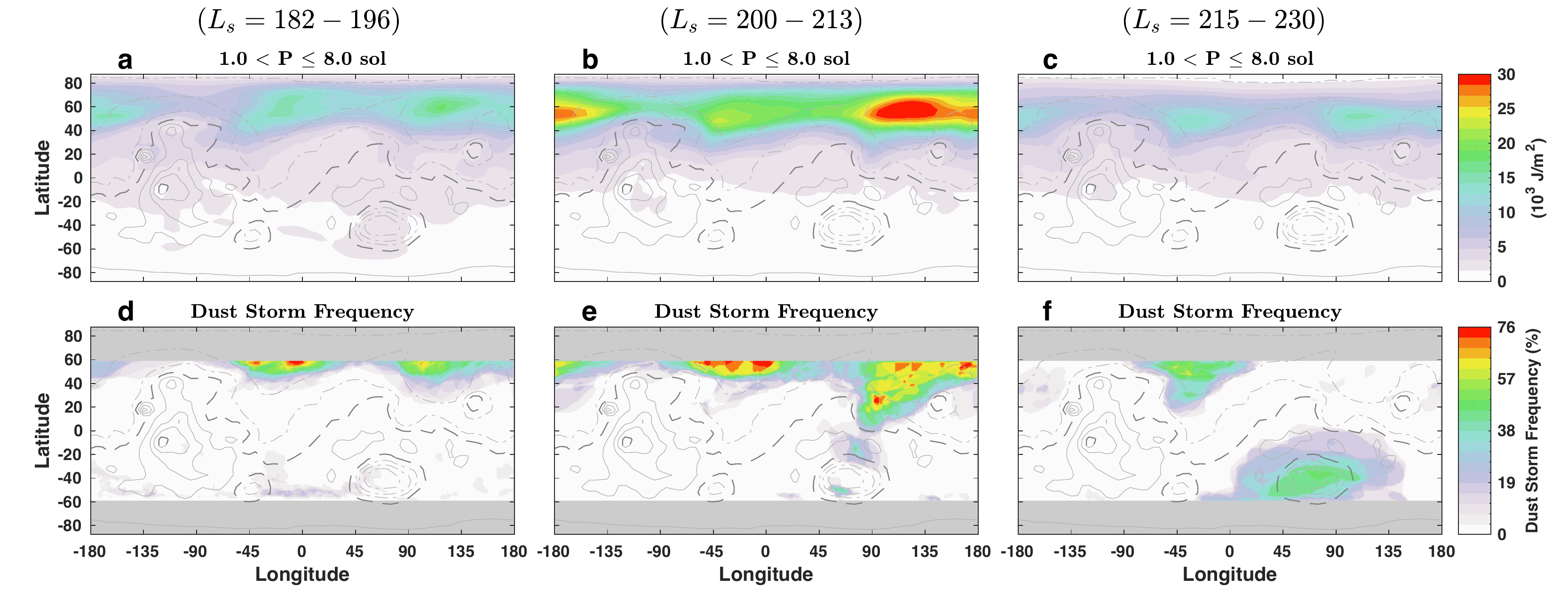}
  \caption{Eddy kinetic energy, $\overline{KE_{t,\lambda,\theta}}$, for $1<P\leq8$ sol eddies during $L_s$ = 182$^{\circ}$ -- 196$^{\circ}$ (panel a), $L_s$ = 200$^{\circ}$ -- 213$^{\circ}$ (panel b), and $L_s$ = 215$^{\circ}$ -- 230$^{\circ}$ (panel c) in MY 26.  The bottom row (panels d--f) shows the dust storm occurrence frequency derived from MDGMs for each corresponding time period.  Topography is indicated by gray contours.}\label{map26a}
\end{figure}

\begin{figure}[tph]
  \noindent\includegraphics[width=29pc,angle=0]{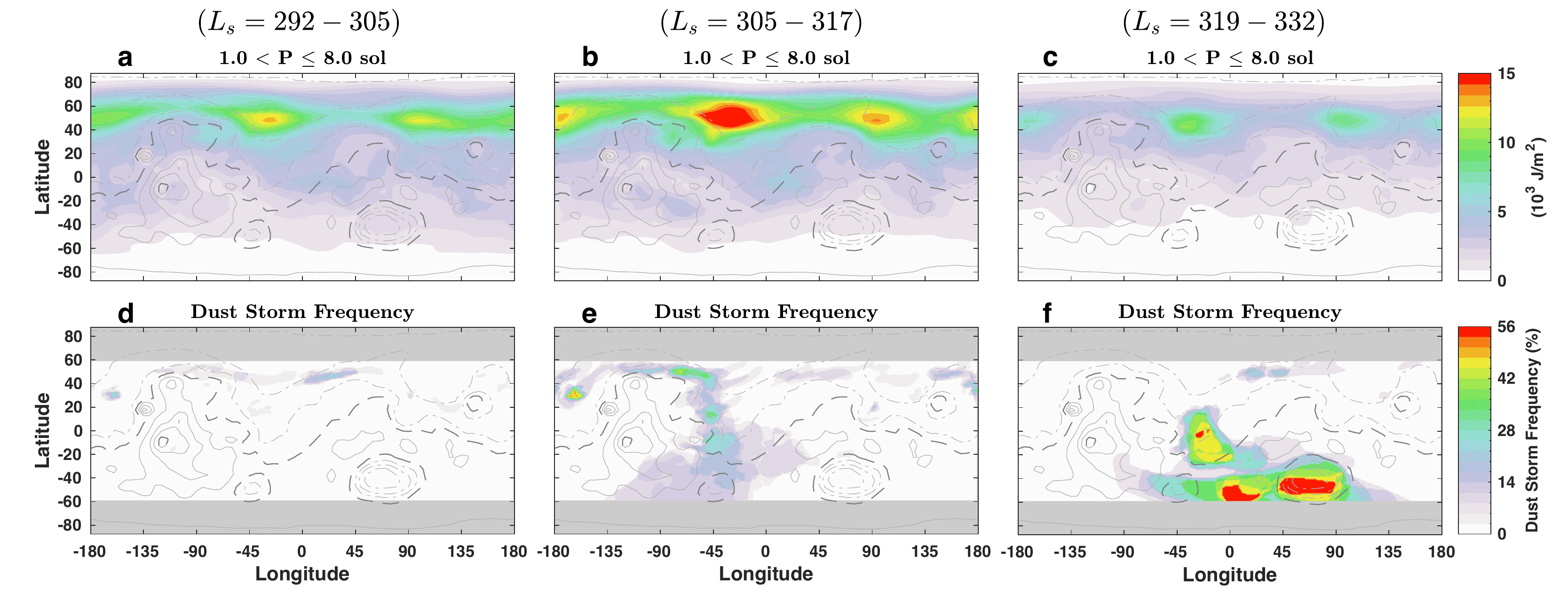}
  \caption{As in Fig. \ref{map26a} but for $L_s$ = 292$^{\circ}$ -- 305$^{\circ}$ (panel a), $L_s$ = 305$^{\circ}$ -- 317$^{\circ}$ (panel b), and $L_s$ = 317$^{\circ}$ -- 322$^{\circ}$ (panel c) in MY 26.}\label{map26c}
\end{figure}

The feedback between the generation of dust activity associated with the $1<P\leq8$ sol eddies and subsequent suppression of the eddies by dust is similar for MY 26, despite the lower confidence in MACDA during MY 26 due to the previously mentioned issue with the TES polar temperature. Figure \ref{map26a} shows the corresponding time periods before, immediately preceding, and after the flushing behavior of the MY 26 ``A'' storm ($L_s$ = 182$^{\circ}$ -- 196$^{\circ}$, $L_s$ = 200$^{\circ}$ -- 213$^{\circ}$, and $L_s$ = 215$^{\circ}$ -- 230$^{\circ}$), and Figure \ref{map26c} shows those periods for the MY 26 ``C'' storm ($L_s$ = 292$^{\circ}$ -- 305$^{\circ}$, $L_s$ = 305$^{\circ}$ -- 317$^{\circ}$, and $L_s$ = 319$^{\circ}$ -- 332$^{\circ}$).  The strongest $\overline{KE_{t,\lambda,\theta}}$ is contained within the Arcadia, Acidalia, and Utopia storm zones during all six periods (Figs. \ref{map26a}a -- c and \ref{map26c}a -- c).  The correspondence between the largest increase of $1<P\leq8$ sol eddies and the primary flushing channel also holds for the MY 26 ``A'' (Fig. \ref{map26a}b) and ``C'' (Fig. \ref{map26c}b) storms.  The MY 26 ``A'' storm occurred primarily through Utopia (Fig. \ref{map26a}e), and the MY 26 ``C'' storm flushed via Acidalia (Fig. \ref{map26c}e).  In both cases, once lifting in the southern hemisphere occured in earnest (Figs. \ref{map26a}f and \ref{map26c}f), the $\overline{KE_{t,\lambda,\theta}}$ in the respective channels dropped back to pre-storm levels (Figs. \ref{map26a}c and \ref{map26c}c).

Returning to Fig. \ref{map} for MY 24, the $8<P\leq60$ sol eddies are increasingly amplified with time from $L_s$ = 195$^{\circ}$ -- 207$^{\circ}$ to $L_s$ = 227$^{\circ}$ -- 240$^{\circ}$ (Fig. \ref{map}c, g, \& k). They acquire noticeable amplitude near Acidalia during $L_s$ = 210$^{\circ}$ -- 223$^{\circ}$ where and when successive flushing events occur. The global average of their $\overline{KE_{t,\lambda,\theta}}$ increases by $\sim$128$\%$ from $L_s$ = 210$^{\circ}$ -- 223$^{\circ}$ (1.28 $\times$ 10$^3$ J/m$^{2}$) to $L_s$ = 227$^{\circ}$ -- 240$^{\circ}$ (2.91 $\times$ 10$^3$ J/m$^{2}$). The eddies occupy the winter (northern) hemisphere, consistent with the spatial distribution of long wave period traveling waves derived from TES data \citep{Banfield2004}. The largest amplitude of the $8<P\leq60$ sol eddies occurs in the broad longitude sector from northwest Tharsis across Acidalia to Utopia. 

Normalizing the results by pressure using Eq. \ref{eq:KEnor} for MY 24 (Supplementary Fig. 4) increases the amplitude of the $\overline{KE_{t,\lambda,\theta}}_{norm}$ of the $P\leq1$ sol eddies relative to the other wave periods due to the lower surface pressures in the southern hemisphere.  However, these changes are minimal, and the temporal variation of each eddy category remains the same as that shown in Fig. \ref{map}. 

 \section{Summary and discussion}\label{summary}

Using the MACDA reanalysis product and the MGS MDGM dust storm database, we have examined the changes in the kinetic energy of various eddies at different development stages of representative large regional dust storms in MY 24 and 26. Similar large regional dust storms typically occur before and after the northern winter solstice  during the dust storm season in non-GDS years. A conceptual model for the evolution of atmospheric eddies during the lifetime of large dust storms can be constructed:  Repeating peaks in $1<P\leq8$  sol eddies trigger one or more flushing dust storm sequences through Acidalia, Utopia, and/or Arcadia storm tracks;  Dust storms propagate southward across the equator and zonally expand in the southern hemisphere, increasing the global-mean dust opacity; The $P\leq1$ sol eddies--composed mostly of the thermally forced tides--sharply amplify toward the late stage of this process and peak before the maximum global dust opacity is reached; Subsequently, the $1<P\leq8$ sol eddies are depressed; and The $8<P\leq60$ sol eddies are prominently amplified during a brief time period when the global mean dust opacity remains high. With the decline of global dust opacity, the circulation returns to its seasonal strength. 


The $P\leq1$ sol eddies are found to be dominant most of the time, are highly sensitive to amount of atmospheric dust opacity, and  peak slightly before the global-mean dust opacity. These eddies are amplified even if dust storm sequences do not lead to a major dust storm. In comparison, the $8<P\leq60$ sol eddies are usually the weakest but show a prominent peak during major dust storms that have large-scale impact. If active flushing dust storms do not result in a major dust storm that significantly elevates the global dust opacity, the $8<P\leq60$ sol eddies remain weak. The $1<P\leq8$ sol eddies accompany dust storm sequences. These synoptic eddies rival (and occasionally exceed) the $P\leq1$ eddies before the global-mean dust opacity is significantly enhanced, and the $1<P\leq8$ sol eddies are greatly suppressed afterwards.

The $1<P\leq8$ sol eddies contain large contributions from zonal wavenumbers $k$ = 1 -- 3. Sometimes, a single wavenumber dominates, but more frequently, combinations of different zonal wavenumbers contribute to the peaks.  These peaks correlate reasonably with textured frontal/flushing dust storms between 40$^\circ$N -- 60$^\circ$N, suggesting that $1<P\leq8$ sol traveling waves can raise and concentrate dust into frontal dust storms.  The $8<P\leq60$ sol and $P\leq1$ sol eddies are each dominated by $k$ = 1.

We divide the MY 24 ``A'' storm into time periods before, during, and after the height of frontal/flushing events ($L_s$ = 195$^{\circ}$ -- 207$^{\circ}$, $L_s$ = 210$^{\circ}$ -- 223$^{\circ}$, and $L_s$ = 227$^{\circ}$ -- 240$^{\circ}$).  The $1<P\leq8$  sol eddies first strengthen, which is consistent with an increased dust storm frequency within the northern hemisphere storm tracks during $L_s$ = 210$^{\circ}$ -- 223$^{\circ}$, particularly in the primary Acidalia flushing track; then, the $1<P\leq8$ sol eddies weaken as increased global-mean dust opacity suppresses baroclinic eddies.  The $8<P\leq60$ sol eddies become increasingly amplified with time across a broad longitudinal sector that includes Acidalia.  Overall, with the development of the MY 24 ``A'' storm, the strongest eddies progress from $1<P\leq8$ sol eddies in northern mid/high latitudes to $P\leq1$ sol eddies in the southern mid/high latitudes. During the process, the $8<P\leq60$ sol eddies strengthen markedly, though they are never dominant among the eddies examined.

Previous work has linked various circulation changes with the occurrence of large dust storms \citep{Zurek1992,Barnes2017}. This paper concentrates on linking the timing and spatial distribution of changes in various eddies with the development stages of pre-solstice ``A'' and post-solstice ``C'' storms. It will be enlightening to study similar events from other Mars years to further understand Martian dust storm dynamics and inter-annual variability in future work. 

\section*{Acknowledgments}
MACDA v1 is available at $<$http://macdap.physics.ox.ac.uk$>$. The gridded optical depth data of \cite{Montabone2015} are available at $<$http://www-mars.lmd. jussieu.fr/mars/dust$\_$climatology/index.html$>$. MGS MOC MDGMs are available at $<$https://dataverse.harvard.edu/dataset.xhtml?persisten tId=doi:10.7 910/DVN/WWRT1V$>$. The analysis in this study is supported by NASA's Mars Data Analysis Program (MDAP) grant 80NSSC17K0475.  We thank two anonymous reviewers for helpful comments.

\section*{References}
\bibliography{references}
\newpage
\section*{Supplementary figures}

\begin{figure}[tph]
  \noindent\includegraphics[width=29pc,angle=0]{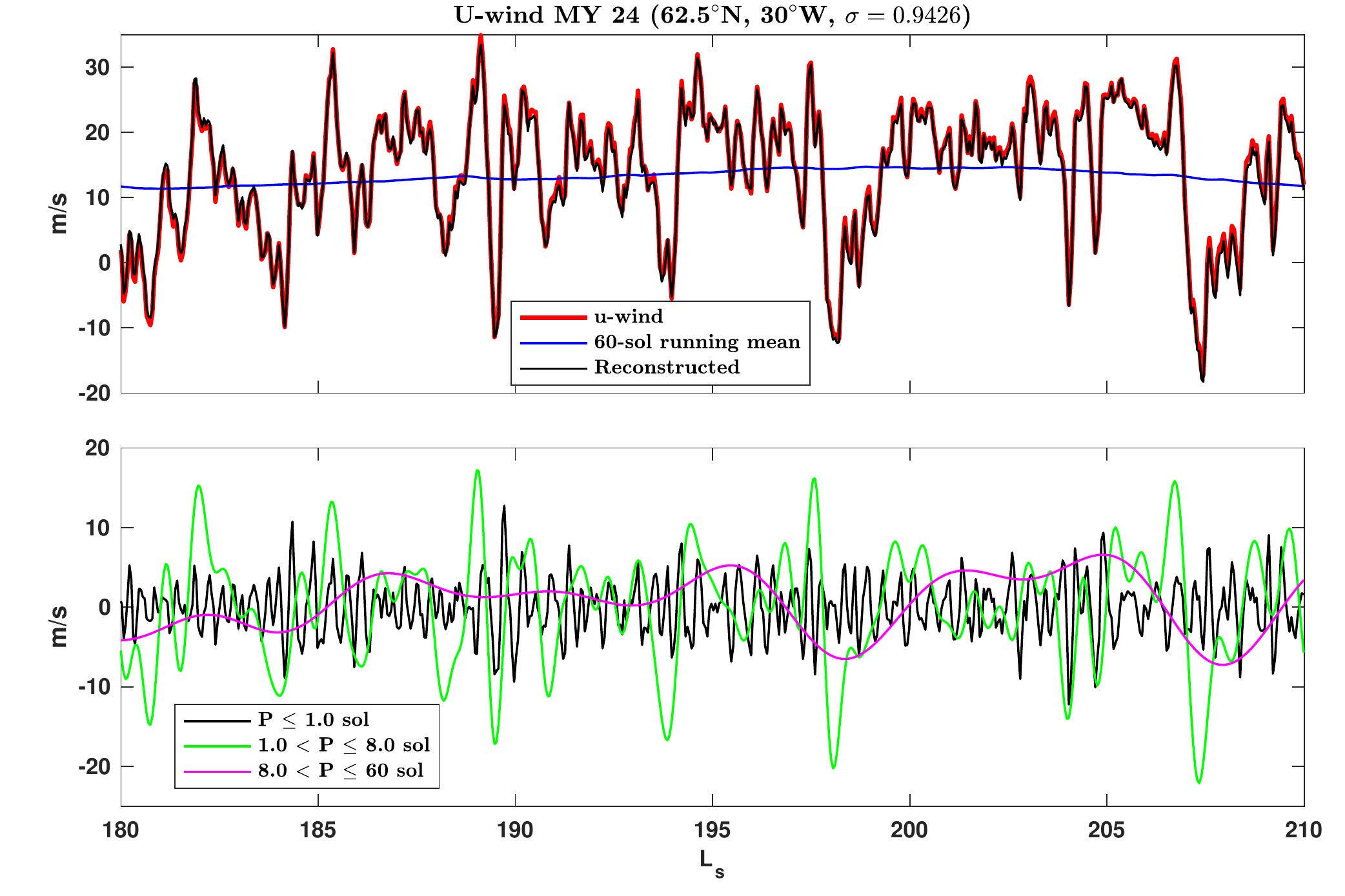}
  \caption*{Supplementary Figure 1:  A segment of zonal wind time series and its eddy components at 62.5$^\circ$N, 30$^\circ$W, and $\sigma$ = 0.9426 ($\sim$600 m) during $L_s$ = 180$^{\circ}$ -- 210$^{\circ}$ in MY 24.  a)  The raw $u$-wind time series from MACDA (red), the 60-sol running mean (blue), and the reconstructed time series (black) using the sum of the filtered eddies from panel b and the 60-sol running mean.  b)  Eddy components of the $u$-wind time series for the $P\leq1$ sol (black), $1<P\leq8$ sol (green), and $8<P\leq60$ sol (magenta) eddies. }
\end{figure}

\begin{figure}[tph]
  \noindent\includegraphics[width=29pc,angle=0]{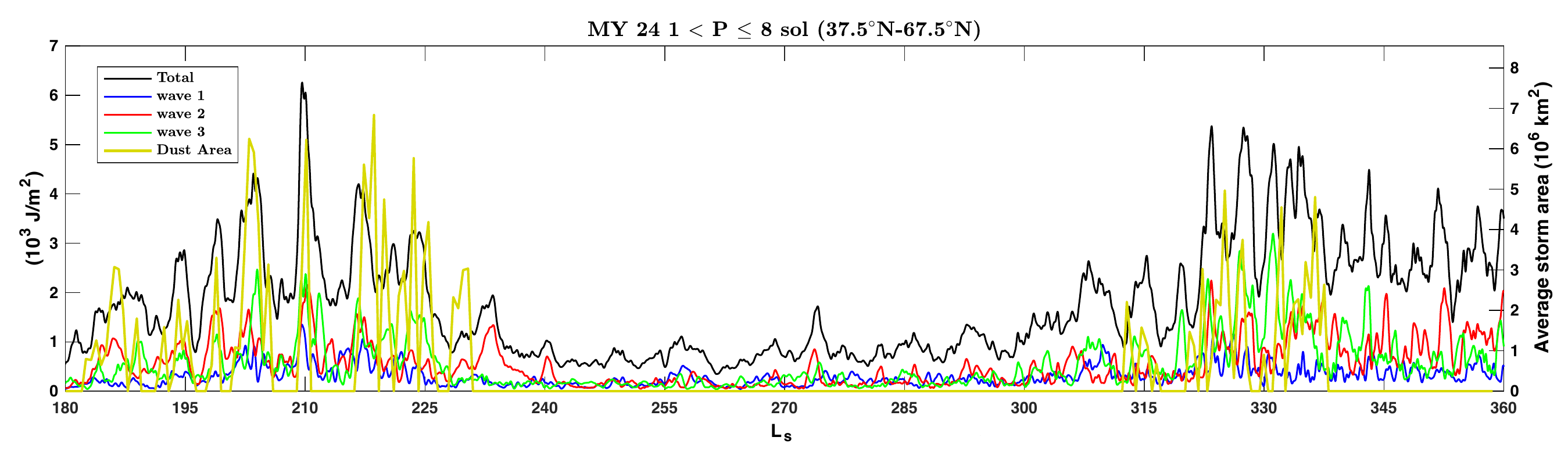}
  \caption*{Supplementary Figure 2:  As in Fig. 2 but  using Equation \ref{eq:KEnor} and \ref{eq:KEt} to calculate the global average of normalized eddy $\overline{KE_{t,\lambda,\theta}}_{norm}$ (J/kg) for (a) and (c).}
\end{figure}

\begin{figure}[tph]
  \noindent\includegraphics[width=29pc,angle=0]{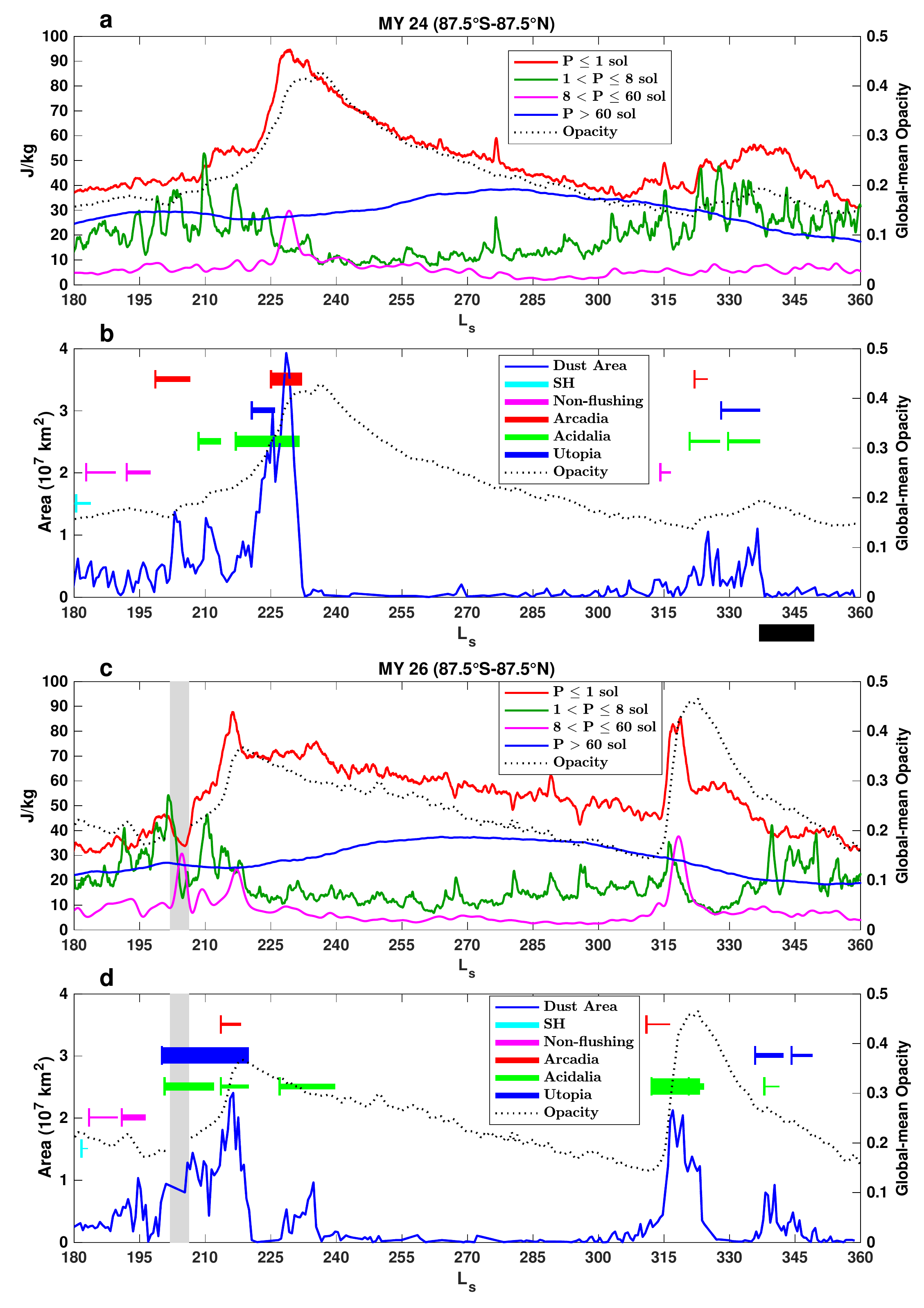}
  \caption*{Supplementary Figure 3:  As in Fig. 3 but the eddy kinetic energy is integrated within $z$ = 0 -- 20 km and 37.5$^\circ$N -- 67.5$^\circ$N for the $1<P\leq8$ sol eddies.}
\end{figure}

\begin{figure}[tph]
  \noindent\includegraphics[width=29pc,angle=0]{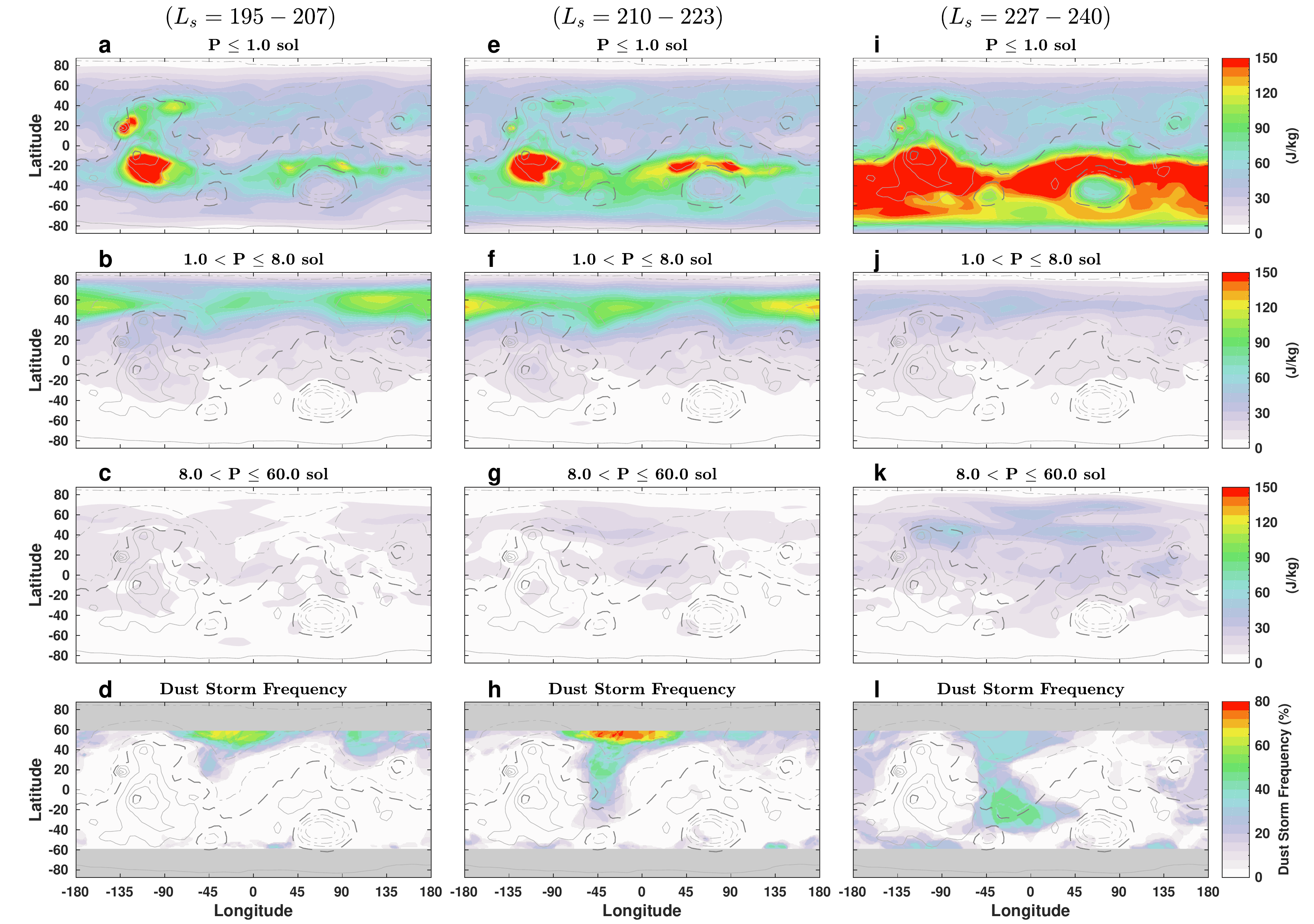}
  \caption*{Supplementary Figure 4:  As in Fig. 5 but using Equation \ref{eq:KEnor} and \ref{eq:KEt} to calculate the global average of normalized eddy $\overline{KE_{t,\lambda,\theta}}_{norm}$ (J/kg).}
\end{figure}

\end{document}